\documentclass[%
reprint,
amsmath,amssymb,
aps,
]{revtex4-2}

\usepackage{graphicx}
\usepackage{dcolumn}
\usepackage{bm}
\usepackage{xcolor}
\usepackage[normalem]{ulem}

\usepackage{subfig}

\begin{document}
	
	\preprint{APS/123-QED}
	
	\title{Null and timelike circular orbits from equivalent 2D metrics}
	
	\author{Pedro V. P. Cunha}
	\author{Carlos A. R. Herdeiro}
	\author{João P. A. Novo}%
	\affiliation{%
		Departamento de Matem\'atica da Universidade de Aveiro \\
		and Centre for Research and Development  in Mathematics and Applications (CIDMA),\\ Campus de Santiago, 3810-183 Aveiro, Portugal
	}%

	\date{July 2022}
	
	\begin{abstract}
		The motion of particles on spherical $1+3$ dimensional spacetimes can,
		under some assumptions, be described by the curves on a 2-dimensional
		manifold, the optical and Jacobi manifolds for null and timelike curves, respectively. In this paper we resort to auxiliary $2$-dimensional metrics to study circular geodesics of generic static, spherically symmetric, and asymptotically flat $1+3$ dimensional spacetimes, whose functions are at least $C^2$ smooth. This is done by studying the Gaussian curvature of the bidimensional equivalent manifold as well as the geodesic curvature of circular paths on these. This study considers both null and timelike circular geodesics. The study of null geodesics through the optical manifold retrieves the known result of the number of light rings (LRs) on the spacetime outside a black hole and on spacetimes with horizonless compact objects. With an equivalent procedure we can formulate a similar theorem on the number of marginally stable timelike circular orbits (TCOs) of a given spacetime satisfying the previously mentioned assumptions.
	\end{abstract}
	
	\maketitle
	
	
	\section{Introduction}
	
	Gibbons and Werner \cite{Gibbons:2008rj} showed how one could use the optical metric of a spherically symmetric spacetime, which is the 2 dimensional Riemannian metric seen by a massless particle, to compute the deflection angle of light by means of the Gauss-Bonnet theorem on the optical manifold. Later, the procedure was generalised to consider the more general axisymmetric spacetimes in two ways, one developed by Werner \cite{Werner:2012rc} which makes use of Nazim's construction of the osculating manifold, and the other developed by Ono et al. \cite{Ono:2017pie} which computes explicitly the contribution of the rotation  1-form to the geodesic curvature of curves and even takes into account finite distance corrections \cite{Ishihara:2016vdc}. The optical metric has been recently widely used to compute the deflection angle of light passing near a compact object both within and beyond GR, $e.g.$ \cite{example1Jusufi:2015laa,example2Ovgun:2018prw,example3Ovgun:2018ran,example4Ovgun:2018fnk,example5Jusufi:2018jof,example6Zhu:2019ura,example7Kumar:2019ohr,Islam:2020xmy}. 
	
	In a procedure similar to the derivation of the optical one can derive the Jacobi metric \cite{Gibbons:2015qja,Chanda:2019guf}, which is also a $ 2 $ dimensional metric but yields the timelike geodesics of the spacetime. Therefore one can use the Jacobi metric to study the paths of massive (and possibly charged) particles in spacetimes \cite{Das:2016opi}. Applying the same reasoning used to compute the deflection angle of light, the Jacobi metric has been used to compute the deflection angle of massive particles passing near compact objects \cite{Crisnejo:2018uyn,Li:2019qyb}.
	Since the Jacobi metric yields the timelike geodesics of the spacetime it can be used, in particular, to study those that are circular, the TCOs.
	
	Unlike LRs, which are null circular orbits, and exist only for some distinct radial coordinates, TCOs can exist for continuous ranges of the radial coordinate. This means that TCOs form a continuum of connected orbits characterised by the energy and angular momentum of the particle, hence unlike LRs the TCOs are not solely characterised by the properties of the spacetime. However, there are special TCOs which depend only on the underlying spacetime, the ones that separate regions of different stability of TCOS. These are known as marginally stable circular orbits (MSCOs), the innermost of which is the innermost stable circular orbit (ISCO). The ISCO, as the name suggests, separates a region where TCOs are stable (radially above in the Schwarzschild case) from another where they are unstable (radially below). For the Kerr metric, regardless of the spin parameter, there is only one such orbit separating regions of stability; however, for other compact objects one can have much richer structure, namely several disconnected regions where stable TCOs are possible \cite{Delgado:2021jxd}. ISCOs are astrophysically relevant due to their impact on the accretion disk dynamics as well as on the study of extreme mass ratio inspirals (EMRIs)  where the motion of the lighter body can be modelled as following stable TCOs, moving gradually inwards, until it reaches the ISCO, after which it plunges towards the central object. Therefore the structure of TCOs and the location of the ISCO greatly affects the emitted gravitational waves (GWs). The GWs generated by such events are expected to lie in the sensitivity range of LISA \cite{Barausse_2020}; hence, the study of such matters is quite timely.
	
	Recently the optical metric has also been used in a discussion of light rings (LRs) on Schwarzschild-like spherically symmetric black holes (BHs) \cite{Qiao:2022jlu}. In the latter it is argued that the Hadamard-Cartan theorem can be used to determine the stability  of the LRs of the spacetime. In the present paper a similar study will be made concerning the TCOs of static, spherically symmetric, asymptotically flat $1+3$ dimensional spacetimes, whose metric functions are at least $C^2$ smooth, using for this purpose the Jacobi metric. We will also extend the analysis to the number of LRs and their stability, similar to what was done in \cite{Cvetic:2016bxi} for BHs. In addition, we also recover the results of recent theorems for stationary spacetimes in the spherically symmetric case~\cite{Cunha:2017qtt,Cunha:2020azh}. However the study of LRs performed here serves to present a novel approach and as a proof of concept before considering TCOs. The equivalent analysis concerning the Jacobi metric will provide similar theorems concerning the number of MSCO on the considered spacetimes.
	
	\section{The spacetime}
	
	In this paper a general spherically symmetric, static, asymptotically
	flat, $1+3$ dimensional metric is considered. It can be described by the following
	line element:
	\begin{equation}
		{\rm d}s^{2}=-f\left(r\right){\rm d}t^{2}+\frac{{\rm d}r^{2}}{h\left(r\right)}+r^{2}{\rm d}\Omega^{2}\,,\label{eq:LineElement}
	\end{equation}
	where ${\rm d}\Omega^{2}$ is the usual metric on the unit 2-sphere and $f,g$ are at least $C^2$ smooth. Asymptotic flatness is imposed by requiring the metric to be Minkowski at infinity, this is achieved by:
	\begin{equation}
		\lim_{r\rightarrow\infty}f\left(r\right)=1\,,\quad\lim_{r\rightarrow\infty}h\left(r\right)=1\,. \label{eq:MinkLimit}
	\end{equation}
	
	We will be concerned with the motion of massless as well as massive
	particles on this spacetime. But we will consider the motion on equivalent bidimensional manifolds, namely the optical and Jacobi manifolds. This allows the usage of several results from differential geometry which can give novel insights into the spacetime geodesics.
	
	Two contrasting spacetime types will be considered: one containing a BH and another describing a horizonless compact object. BHs are characterised by the presence of an event horizon, which for metrics of the form of Eq. (\ref{eq:LineElement}) occurs at $ r=r_H $ such that (check App. \ref{horizons})
	\begin{equation}
		h\left(r_H\right)=0\,.
	\end{equation}
	Carter proved that for static BHs the event horizon must always be also a Killing horizon of the time translation Killing vector field $\partial_{t}$ \cite{Carter:1973rla, Wald:1999vt}. In the present case, this implies that $f\left(r_{H}\right)=0$. This is a purely geometric result that does not invoke Einstein's field equations. This Killing vector field must remain timelike everywhere outside the horizon, such that $f\left(r\right)>0\,,r>r_{H}$ (see section 12.3 of \cite{Wald:1984rg}).
	
	 Only non-extremal BHs will be considered. Extremal BHs are defined by the vanishing of their surface gravity, $ \kappa_{\mathrm{surf}} $. For the spacetimes considered the surface gravity is given by:
	\begin{equation}
		\kappa_{\mathrm{surf}}=\lim_{r\rightarrow r_{H}}\sqrt{\frac{h\left(r\right)}{f\left(r\right)}}\frac{f^{\prime}\left(r\right)}{2}\,.
	\end{equation}
	For spherically symmetric BHs this corresponds to the acceleration of a static observer at the horizon as measured at spatial infinity. The surface gravity  of non-extremal BHs satisfies $ \kappa_{\mathrm{surf}}>0 $, the derivation and a small discussion is found in App. \ref{sec:SurfGravity}.
	
	Like the name suggest horizonless compact objects have no horizon therefore  $ f\left(r\right)>0 $ and $ h\left(r\right)>0 $ everywhere, for line elements of the form (\ref{eq:LineElement}). The spacetime should be regular, namely at $ r=0 $; then, the functions $ f $ and $ h $ can be Taylor expanded around $ r=0 $. Inserting the resulting expansions on the expressions for the Ricci and Kretschmann scalars one finds that for them to be regular at $ r=0 $ one must have as $ r\rightarrow 0 $ \cite{Brito:2015pxa}:
	\begin{equation}
		f\left(r\right)= f_{0}+f_{2}r^{2}+\mathcal{O}\left( r^4\right) \,,\quad
		h\left(r\right)= 1+h_{2}r^{2}+\mathcal{O}\left( r^4\right)\,. \label{eq:OriginHCO}
	\end{equation}
	
	In this paper topologically non-trivial spacetimes will not be considered, such as wormhole spacetimes.
	\section{Light rings}
	
	This section will focus on the circular null curves of the spacetime, which when these are geodesics correspond to the LRs. This study will be carried using the optical metric, to be defined below. This formalism will be applied to two distinct compact objects spacetimes, those with horizon (BHs) and those without horizons. Asymptotic flatness of the spacetime implies the same behaviour of the metric at infinity, however in the origin/horizon limit the behaviour will be distinct for the two cases.
	
	\subsection{The optical metric}
	
	The optical metric is the metric as seen by a massless particle. It
	is obtained by inserting the null condition, ${\rm d}s^{2}=0$, into
	the line element (\ref{eq:LineElement}) and solving for ${\rm d}t$.
	This yields
	\begin{equation}
		{\rm d}t^{2}=\frac{1}{f\left(r\right)}\left(\frac{{\rm d}r^{2}}{h\left(r\right)}+r^{2}{\rm d}\Omega^{2}\right)=\frac{1}{f\left(r\right)}\left(\frac{{\rm d}r^{2}}{h\left(r\right)}+r^{2}{\rm d}\phi^{2}\right)\,.\label{eq:OpticalGeneric}
	\end{equation}
	For the last equality we took advantage of the symmetry of the problem
	and restricted the analysis to the equatorial plane of the optical
	manifold, $\theta=\pi/2$. The geodesics of this metric are the light rays, which are defined as the spatial projections of the null geodesics of the original spacetime, Eq. (\ref{eq:LineElement}).
	This formulation can be used to compute the deflection angle of light
	for several spacetimes by means of the Gauss Bonnet theorem. However
	for this paper we will consider only the Gaussian curvature of the
	manifold as well as the geodesic curvature of circular orbits. These
	are given by, respectively,
	\begin{align}
		K & =-h^{\prime}(r)\frac{2f(r)-rf^{\prime}(r)}{4r}\nonumber\\
		& \quad+\frac{h(r)}{2}\left[f^{\prime}(r)\left(\frac{1}{r}-\frac{f^{\prime}(r)}{f(r)}\right)+f^{\prime\prime}(r)\right]\label{eq:GaussLR}\\
		\kappa_g & =\sqrt{\frac{h\left(r\right)}{f\left(r\right)}}\frac{2f(r)-rf^{\prime}(r)}{2r}\,.\label{eq:kappaLR}
	\end{align}
	The full computation can be found in App. \ref{sec:Surfaces}. By definition, the geodesic curvature vanishes for geodesics, hence the roots of
	$\kappa_g\left(r\right)=0$ correspond to null circular geodesics of the spacetime, $i.e.$ the LRs.
	We can make use of this feature to confirm several theorems concerning the number and stability of LRs on a spacetime of the form (\ref{eq:LineElement}), by studying the asymptotic behaviour of this function. For now only the simplest case will be considered: the zeros of $\kappa_g$ are assumed simple, $i.e.$ points where $\kappa_g=0$ and $\kappa_g^\prime\neq0$. In doing so degenerate LRs, corresponding to the coalescence of two LRs, are avoided. A brief comment on the degenerate case  will be made at the end of this section. 
	
	\subsection{Asymptotic limit}
	
	Inserting the conditions for asymptotic flatness, Eq. (\ref{eq:MinkLimit}), into Eq. (\ref{eq:kappaLR}) one obtains
	\begin{equation}
		\lim_{r\rightarrow\infty}\kappa_g=\frac{1}{r}\,.
	\end{equation}
	This means that $\kappa_g$ tends asymptotically to zero from positive
	values. Such result is easily understood. Spacetime asymptotic flatness
	 implies that the optical metric, Eq. (\ref{eq:OpticalGeneric}),
	is asymptotically Euclidean and $1/r$ is precisely the curvature
	of a circumference in Euclidean space.
	
	\subsection{Black holes - horizon limit}
	
	Having considered the asymptotic limit of the geodesic curvature we have to compute
	the opposite limit, which, for spacetimes containing BHs, corresponds to the event horizon.
	
	Considering the horizon limit in Eq. (\ref{eq:kappaLR}) one obtains
	\begin{align}
		\lim_{r\rightarrow r_{H}}\kappa_g=&\lim_{r\rightarrow r_{H}}\sqrt{\frac{h\left(r\right)}{f\left(r\right)}}\frac{2f(r)-rf^{\prime}(r)}{2r}\\
		=&\lim_{r\rightarrow r_{H}}-\sqrt{\frac{h\left(r\right)}{f\left(r\right)}}\frac{f^{\prime}(r)}{2}=-\kappa_{\mathrm{surf}}<0\,.
	\end{align}
	This, together with $\kappa_g\left(+\infty\right)=0^{+}$, implies that
	for non-extremal BH spacetimes $\kappa_g$ must vanish at least once between the horizon and infinity. Moreover, the number of these zeros, which correspond to LRs of the spacetime, must always be odd. This is in agreement with the general theorem in~\cite{Cunha:2020azh}.

	\subsection{Horizonless compact objects - origin limit}
	
	For spacetimes harboring horizonless compact objects, we must consider the limit $ r\rightarrow0 $. Following the expansions presented in Eq. (\ref{eq:OriginHCO}) one obtains for the equatorial curvature of circular null geodesics
	\begin{equation}
		\lim_{r\rightarrow0}\kappa_g=\lim_{r\rightarrow0}\frac{1}{r}=+\infty\,.
	\end{equation}
	This result together with the asymptotic behaviour of $\kappa_g$ indicates that a horizonless compact object does not need to have LRs, and if	it does they will always come in pairs. This is consistent with the general theorem in~\cite{Cunha:2017qtt}.
	
	\subsection{Stability of the LRs}
	
	So far we have discussed only the existence and number of LRs but using the Gaussian curvature of the manifold we can compute their stability. The stability of the LRs are determined by the sign of $K$ at that radius, stable (unstable) orbits have a positive 	(negative) Gaussian curvature. This is discussed in Appendix \ref{sec:Surfaces}. Considering the Gaussian curvature, Eq. (\ref{eq:GaussLR}), at a LR, $i.e.$ at a radial coordinate $r=r_{\mathrm{LR}}$ such that $\kappa_g\left(r=r_{\mathrm{LR}}\right)=0$, one obtains
	\begin{align}
		\left.K\right|_{r=r_{\mathrm{LR}}}=&\left.\frac{h\left(r \right)}{2}\left(f^{\prime\prime}\left(r \right)-\frac{f^{\prime}\left(r \right)}{r}\right)\right|_{r=r_h}\nonumber\\
		=&-\sqrt{h\left(r \right)f\left(r \right)}\left.\kappa_g^{\prime}\right|_{r=r_{\mathrm{LR}}}\,.
	\end{align}
	From this relation it follows that the stability of a LR is determined by the way $\kappa_g$ crosses the horizontal axis at $r=r_{\mathrm{LR}}$,
	if it crosses with positive (negative) slope the associated LR is unstable (stable).
	
	\subsection{Some remarks}
	
	This discussion shows that asymptotically flat spacetimes containing a non-extremal BH have $ 2n+1,\,n\in \mathbb{N}_0 $ LRs, the first and last of which will be unstable, and they will always alternate. Spacetimes describing horizonless compact objects, on the other hand, have $2n,\,n\in\mathbb{N}_{0}$ LRs, where the first (innermost) LR will be stable and the last (outermost) unstable, and they will also always alternate.
	
	It also follows that spacetimes with more than one LR will always have at least one stable LR. The presence of the latter has been argued to imply the spacetime is unstable \cite{Keir:2014oka,Cardoso:2014sna}; in fact this was recently shown in specific examples~\cite{Cunha:2022gde}. Moreover, it has been conjectured that the appearance of more than one LR in BH spacetimes requires the violation of the strong energy condition \cite{Cvetic:2016bxi}.
	
	In this analysis we did not considered the hypothesis of a degenerate root, this would correspond to the case where a zero of $\kappa_g$ corresponds to a maximum or minimum. Such points will have $\kappa^\prime=0\Rightarrow K=0$, hence they will be marginally stable, $i.e.$ stable against perturbations in a given direction and unstable against perturbations in the opposite one. Therefore one should compute the derivative of the Gaussian curvature at these points, $r=\bar{r}$, where $\kappa_g\left(\bar{r}\right)=0=\kappa_g^\prime\left(\bar{r}\right)$. This is given by
	\begin{equation}
	    \left.K^\prime\right|_{r=\bar{r}}=\left.-\sqrt{\frac{f\left(r\right)}{h\left(r\right)}}\kappa_g^{\prime\prime}\right|_{r=\bar{r}}\,.
	\end{equation}
	Hence, if $\kappa_g^{\prime\prime}>0$ the LR will be stable against perturbations that increase the radius and unstable otherwise. If $\kappa_g^{\prime\prime}<0$ the converse happens. This further analysis is merely the first level of additional complexity and does not exhaust all the possibilities; it may occur that several of the derivatives of $\kappa_g$ vanish at some point, and this point is an inflection point if the first non-vanishing derivative is even and an extremum if it is odd. Therefore a full analysis should take this into account. This is however beyond the scope of our discussion; moreover it requires very specific conditions, rather than generic cases. 
    \begin{figure}[h!]
    \subfloat[Horizonless compact objects]{\includegraphics[width=0.45
    \textwidth]{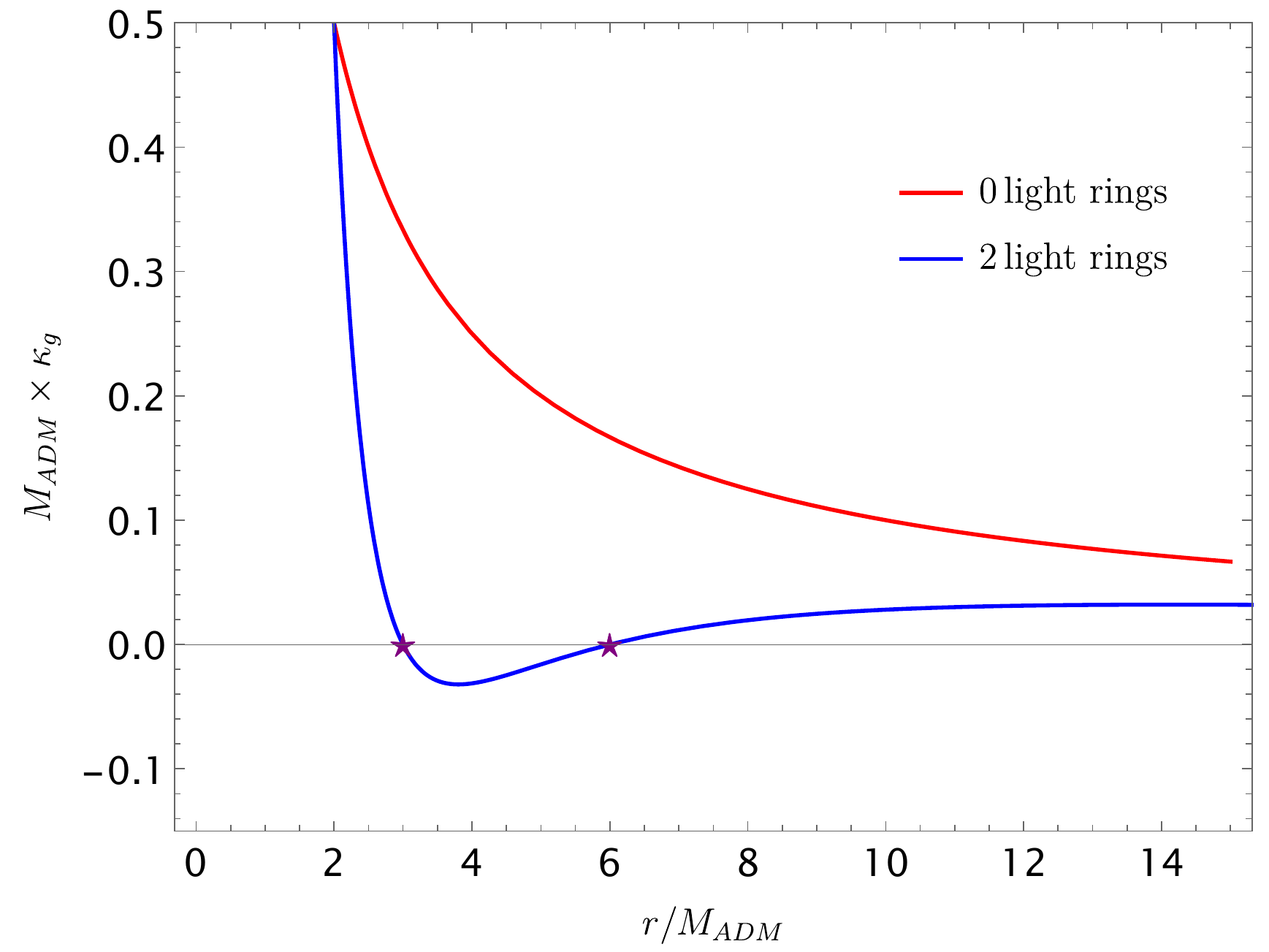}   }
%
\par
\subfloat[Black holes]{\includegraphics[width=0.45\textwidth]{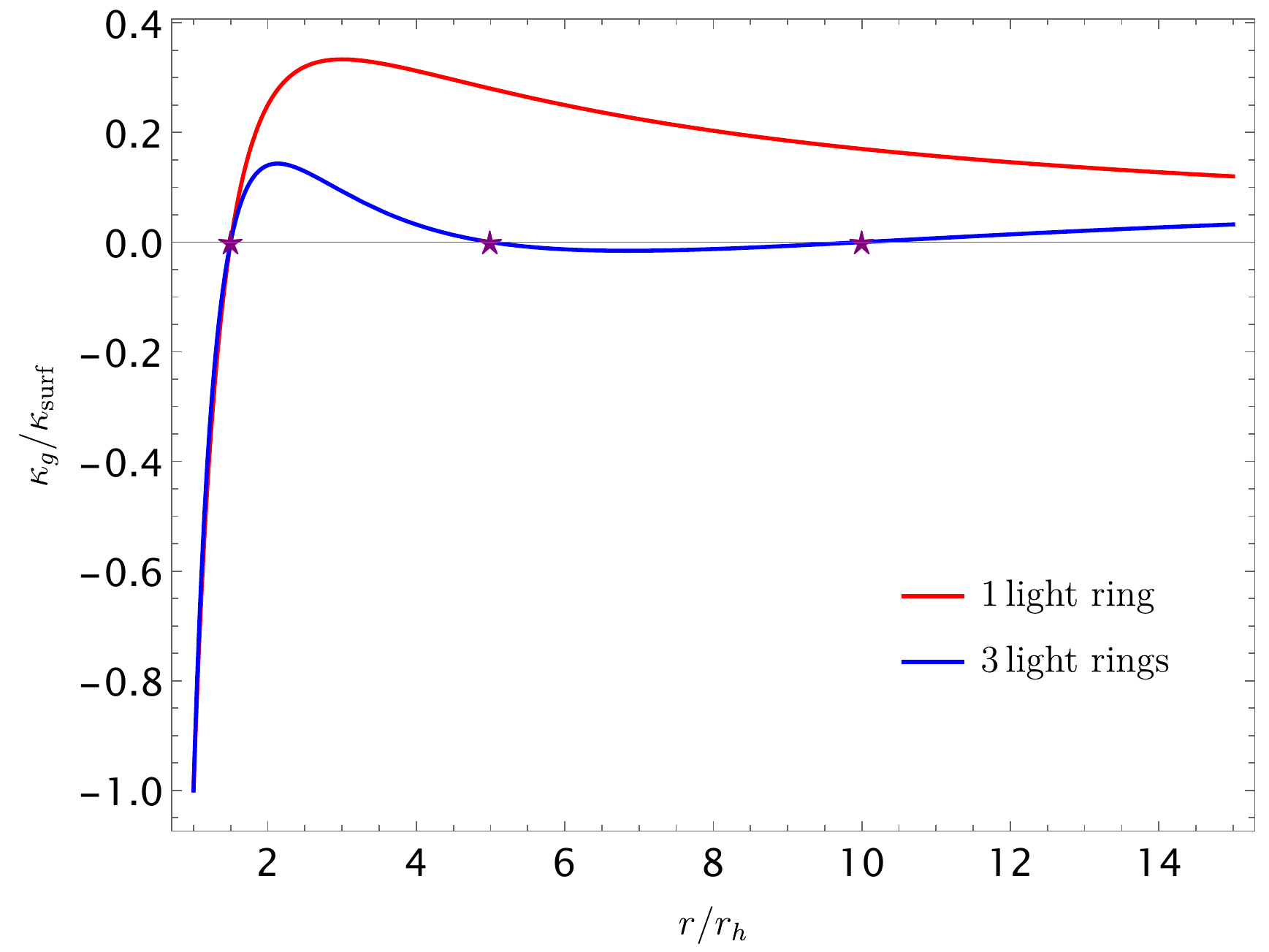}

}
\par
\caption{ Two illustrative plots of the behaviour of $\kappa_g$ for spacetimes containing horizonless compact objects, (a), or BHs, (b). In each plot are represented examples with the lowest and second lowest number of LRs. The LRs are indicated with purple stars.}
\end{figure}
	
	\section{Marginally stable circular orbits}
	
	In this section a similar analysis to the one performed in the optical manifold will be performed in the Jacobi manifold. The goal of the analysis will be to obtain the MSCO of the spacetime, and thus the structure of TCOs on it.
	
	\subsection{The Jacobi metric}
	
	Much like for null particles, the motion of massive ones is also given by the geodesics on a 2 dimensional Riemannian manifold, the Jacobi metric. This metric depends on the energy of the particle and in the massless limit reduces to the optical metric. To obtain the Jacobi metric we recall that test particles of mass $ m $ will move along the timelike geodesics of the spacetime. These can be obtained by extremising the following action
	\begin{equation}
		S=\int{\cal L}{\rm d}\lambda=-m\int\sqrt{-g_{\mu\nu}\dot{x}^{\nu}\dot{x}^{\mu}}{\rm d}\lambda\,,
	\end{equation}
	where $\lambda$ is an arbitrary parameter along the geodesic and the dot denotes differentiation
	with respect to it. From now on it will be assumed that trajectories are parametrised by the coordinate time $t$, such that:
	\begin{equation}
		\mathcal{L}=-m\sqrt{f\left(r\right)-g_{ij}\dot{x}^i \dot{x}^j }\,.
	\end{equation}
	The associate canonical momentum, $ p_i $, is defined in the usual way
	\begin{equation}
		p_i=\frac{\partial\mathcal{L}}{\partial \dot{x}^i}=m\frac{g_{ij} \dot{x}^j}{\sqrt{f\left(r\right)-g_{mn}\dot{x}^m\dot{x}^n}}\,.
	\end{equation}
	With this it is possible to obtain corresponding Hamiltonian, $ H $, as
	\begin{align}
		H&=p_i \dot{x}^i -\mathcal{L}=\frac{m f\left(r\right) }{\sqrt{f\left(r\right)-g_{mn}\dot{x}^m\dot{x}^n}}\nonumber\\
		&=\sqrt{m^2 f\left(r\right)^2+ f\left(r\right)^2 g^{ij}p_i p_j }\,.
	\end{align}
	In the Hamilton-Jacobi formalism the momenta are the gradient of the action function, $ S $,
	\begin{equation}
		p_i=\partial_i S\,,
	\end{equation}
	such that the Hamilton-Jacobi equation reads
	\begin{equation}
		H\left(x^i,\partial_i S, t\right)=-\frac{\partial S}{\partial t}\,.
	\end{equation}
	Since the Hamiltonian does not depend on the parameter $ t $, the right hand side cannot depend on $ t $. Therefore we choose $ -\partial_t S=E $, where $E$ is a constant and can be interpreted as the energy of the system. This choice is consistent with the classical result that for conservative systems the Hamiltonian is time independent and corresponds to the total energy of the system. Rearranging then the Hamilton-Jacobi equation one obtains
	\begin{equation}
	    \alpha^{ij}p_i p_j =1\,,\quad
	    \alpha^{ij}\equiv \frac{f\left(r\right)}{E^2 - m^2 f\left(r\right)}g^{ij} \ .
	\end{equation}
	This is precisely the equation for the geodesics of the Jacobi metric, $ \alpha_{ij} $, defined as
	\begin{equation}
		{\rm d}s_{J}^{2}=\alpha_{ij}\mathrm{d}x^i \mathrm{d}x^j  =\frac{E^{2}-m^{2}f\left(r\right)}{f\left(r\right)}\left[\frac{{\rm d}r^{2}}{h\left(r\right)}+r^{2}{\rm d}\phi^{2}\right]\,.
	\end{equation}
	This expression makes clear that in the massless limit we recover the optical metric, Eq. (\ref{eq:OpticalGeneric}), apart from a conformal factor $ E^2 $, which does not affect the geodesics.
	
	 If a given circular orbit is a geodesic  its geodesic curvature vanishes, which is given by
	\begin{equation}
		\kappa_g=\frac{2f(r)\left(\varepsilon^{2}-f(r)\right)-r\varepsilon^{2}f^{\prime}(r)}{2mr\sqrt{f(r)h(r)}\left(\varepsilon^{2}-f(r)\right)^{3/2}}\,.
	\end{equation}
	where $\varepsilon=E/m$. Much like with the optical metric the radial stability of such orbits is determined by the sign of the Gaussian curvature along it. The Gaussian curvature of this manifold is 
	\begin{widetext}
		\begin{align}
			K= & -\frac{1}{4m^{2}rf(r)\left(f(r)+\varepsilon^{2}\right)^{3}}\left[\varepsilon^{2}f(r)^{2}\left(h'(r)\left(rf'(r)+2\varepsilon^{2}\right)+2h(r)\left(f'(r)+rf''(r)\right)\right)+2r\varepsilon^{4}h(r)f'(r)^{2}\right.\nonumber \\
			& \left.-4\varepsilon^{2}f(r)^{3}h'(r)+2f(r)^{4}h'(r)+\varepsilon^{2}f(r)\left(-f'(r)\left(2h(r)\left(2rf'(r)+\varepsilon^{2}\right)+r\epsilon^{2}g'(r)\right)-2r\varepsilon^{2}h(r)f''(r)\right)\right]\,. \label{eq:GaussTCOs}
		\end{align}
	\end{widetext}

	It should be noted that since these formulas depend only on the square of the energy these equations are also valid for circular geodesics with purely imaginary energies, $\varepsilon^{2}<0$. These will correspond to circular spacelike orbits.
	
	It will be useful later on to introduce the energy of a circular geodesic at a radius $ r $, obtained from solving $ \kappa_g=0 $ for $ \varepsilon $; this yields:
	\begin{equation}
		\varepsilon^2=-\frac{2f\left(r\right)}{2f\left(r\right)-rf^\prime\left(r\right)}\,.\label{eq:EnergyTCOs}
	\end{equation}
	As discussed previously the numerator of this expression is positive everywhere outside the horizon. However the denominator can take on either sign and even be zero, possible zeros would correspond to LRs. Since the left hand side is the square of the energy the right hand side must be positive for real energies. However the right hand side can become negative yielding purely imaginary energies, which correspond to spacelike geodesics. Therefore TCOs can only occur in regions where $ rf^\prime\left(r\right)-2f\left(r\right)>0 $. This corresponds to the region between an unstable LR (at $ r=r_1 $) and a stable one (at $ r=r_2 $) with $ r_2>r_1 $, the region from the outermost LR  (unstable) to infinity, or in the case of horizonless compact objects the region from $ r=0 $ up to the first LR (stable). This is in agreeement with the results found in \cite{Delgado:2021jxd}.
	
	Inserting Eq. (\ref{eq:EnergyTCOs}) into Eq. (\ref{eq:GaussTCOs}) one obtains the Gaussian curvature  along circular geodesics of a given $ r $, this is denoted by $ K^\mathrm{circ} $ and takes the following form
	\begin{widetext}
		\begin{equation}
			K^{\mathrm{circ}}=\frac{h(r)}{m^{2}r^{3}f(r)f'(r)^{2}}\underbrace{\left[2f(r)-rf'(r)\right]}_{\text{Light Rings}}\underbrace{\left[f(r)\left(3f'(r)+rf''(r)\right)-2rf'(r)^{2}\right]}_{\text{MSCOs}}\,.\label{eq:GaussMSCO}
	\end{equation}\end{widetext}
	
	Since the stability is determined by the sign of this expression the MSCOs occur at radius where $ K^\mathrm{circ} $ vanishes. The numerator of this expression is a product of three different expressions, the first from left to right is $ h\left(r\right) $ which is always positive outside the horizon, the zeros of second one correspond to the LRs, Eq. (\ref{eq:kappaLR}), and orbits at these radii have infinite energy per unit mass, this means that the MSCOs will be determined by the final term
	\begin{equation}
		f(r)\left(3f'(r)+rf''(r)\right)-2rf'(r)^{2}=0\,.
	\end{equation}
	
	One should also take into account possible divergences arising from the vanishing of the denominator. From the previous discussion the only points at which this could happen are the extrema of $f$, $i.e$ some $\tilde{r}$ such that $ f^\prime(\tilde{r})=0 $. However since the denominator depends on the square of $f^\prime$ it will always be non-negative, hence $K^\mathrm{circ}$ will have the same sign at $\tilde{r}\pm\delta\,,\delta\ll1$. Since we are only interested in the sign of $ K^\mathrm{circ} $ such poles do not affect the present analysis. The equivalence between this approach, and the usual one with the effective potential is shown in App. \ref{sec:AppPotentialApproach}.
	
	The fact that Eq. (\ref{eq:GaussMSCO}) also yields the LRs justifies the previous analysis of the optical metric, to identify the meaning of that term. This is however somewhat undesirable, so one should study more about the behaviour of $ K^\mathrm{circ} $ at the LRs, namely its slope at these points:
	\begin{equation}
		\left.\frac{\partial K^{\mathrm{circ}}}{\partial r}\right|_{r=r_{\mathrm{LR}}}=\left.-\frac{h(r)\left[r^{2}f^{\prime\prime}(r)-2f(r)\right]^{2}}{4m^{2}r^{3}f(r)^{2}}\right|_{r=r_{\mathrm{LR}}}<0\,.
	\end{equation}
	Therefore $ K^\mathrm{circ} $ crosses the horizon axis at the LRs with negative slope, this means that it must always cross with positive slope between the LRs. If TCOs are possible in that region those crossing points will be the MSCOs and if TCOs are not possible these are marginally stable spacelike orbits.

	\subsection{Asymptotic limit}
	As before we will be concerned with the number of zeros of Eq. (\ref{eq:GaussMSCO}), so we must study its behaviour in the limits of the region where it is defined. The first limit we consider is spatial infinity. Since the spacetime is asymptotically flat the associated Jacobian metric is asymptotically Euclidean, therefore 
	\begin{equation}
		\lim_{r\rightarrow \infty}K^\mathrm{circ}=0\,.
	\end{equation}
	The derivative of $ f $ in the denominator of Eq. (\ref{eq:GaussMSCO}) means that to obtain the behaviour at infinity one needs also to specify how this function decays. It is expected that near infinity the orbits of the spacetime are essentially Keplerian ($e.g.$ the orbits around the supermassive BH at the centre of the galaxy). These orbits are stable, hence $ K^\mathrm{circ} $ must go to zero from above. This is equivalent to requiring that the considered spacetimes reduce to Newtonian gravity near infinity, corresponding to
	\begin{equation}
		f\left(r\right)=1-\frac{2M}{r}+\mathcal{O}\left(r^{-2}\right)\,,\quad h\left(r\right)=1+\mathcal{O}\left(r\right)\,.
	\end{equation}
	Under this assumption one obtains
	\begin{equation}
		\lim_{r\rightarrow\infty}K^{\mathrm{circ}}=\frac{2}{m^{2}Mr}\,.
	\end{equation}
	which yields the expected behaviour.
	
	\subsection{Black holes - horizon limit}
	We are interested in the behaviour of Eq. (\ref{eq:GaussMSCO}) outside the horizon. So now we will consider the horizon limit. As before the horizon is defined by $ h\left(r_H\right)=0=f\left(r_H\right) $, then
	\begin{align}
		\lim_{r\rightarrow r_{H}}K^{\mathrm{circ}}&=\lim_{r\rightarrow r_{H}}\frac{2 h\left(r\right)r^2 f^{\prime}\left(r\right)^3}{m^2 r^3 f\left(r\right)f^\prime\left(r\right)^2}\nonumber\\
		&=\frac{2}{m^{2}r_{H}}\kappa_{\mathrm{surf}}^{2}f^{\prime}\left(r_{H}\right)>0\,.
	\end{align}
	Once again, non-extremality of the BHs is being assumed.

	\subsection{Horizonless compact objects - origin limit}
	For horizonless compact objects, we require regularity of the spacetime at the origin, this implies the same behaviour near the origin expressed in Eq. (\ref{eq:OriginHCO}). This leads to
	\begin{equation}
		\lim_{r\rightarrow0}K^\mathrm{circ}=\frac{4f_0}{m^2r^4f_2}\,.
	\end{equation}
	From the condition that $ f>0 $ for every $ r $ it comes that $ f_0>0 $, however the sign of $ f_2 $ is not constrained, and the divergence will depend on it. For $ f_2\gtrless0 $ one has $ \lim_{r\rightarrow0}K^\mathrm{circ}=\pm\infty $. To explore the physical meaning of $f_2$ one studies the Ricci tensor at the origin, to find
	\begin{equation}
	    \lim_{r\rightarrow0}R_{00}=3f_2\,.
	\end{equation}
	Raychaudhuri's equation states that in order for gravity to be attractive one must have $R_{00}\geq0$. This is the Strong Energy Condition:
	\begin{equation}
	    R_{\mu\nu}t^\mu t^\nu\geq0\,,
	\end{equation}
	where $t^\mu$ is any timelike vector field \cite{Carroll:2004st}.
	
	\subsection{A theorem on the number and location of MSCOs}
	
	Considering first BHs, it was seen that $ K^\mathrm{circ}>0 $ both at the horizon and at spatial infinity. At first glance this seems to indicate that BHs do not necessarily possess ISCOs, contradicting previous results. However as seen before BHs always possess at least one unstable LR, at which $ K^\mathrm{circ} $ vanishes with negative slope. The same behaviour occurs at every LR. As discussed above TCOs may occur in a region from an unstable to a stable LR, which means that in that region the spacetime will have $ 2n+1\,$ MSCOs with $n\in \mathbb{N}_0 $. The same result applies from the outermost LR (which is unstable) up to spatial infinity.\\
	
	In the case of horizonless compact objects, the behaviour of $ K^\mathrm{circ}>0 $  will depend on the sign of    $ f^{\prime\prime}\left(0\right)=f_2 $. The study of these spacetimes can be divided into three distinct regions:
	\begin{itemize}
		\item[$ (i) $] between $ r=0 $ and the innermost LR (which is stable), or spatial infinity if no LRs are present;
		\item[$ (ii) $] between a stable and an unstable LR;
		\item[$ (iii) $] from the outermost LR (which is unstable) to spatial infinity.
	\end{itemize}	
	The sign of $ f_2 $ affects the behaviour near the origin, hence it will only affect region $ (i) $, if $ f_2>0 $ one has $ 2n $ MSCOs and if $ f_2<0 $ one has $ 2n+1 $ MSCOs with $ n\in\mathbb{N}_0 $. In regions $ (ii) $ and $ (iii) $ there must be at least one MSCO, and if more exist they must come in pairs, meaning $ 2n-1 $ MSCOs. Recall that this discussion assumes non degenerate MSCOs.

	In the regions where TCOs are not possible, i.e. between a stable and an unstable LR the zeros of $K^{\mathrm{circ}}$ correspond to marginally stable circular spacelike orbits, of which there is always an odd number.\\
	
	For a recent and complementary discussion on the number of TCOs using a topological approach, in a stationary BH spacetime, see~\cite{Wei:2022mzv}.
	
	\section{Discussion and final remarks}
	
	This paper considers equatorial circular geodesics in arbitrary static, spherically symmetric, $1+3$ dimensional spacetimes that are $C^2$ smooth and asymptotically flat, with a main focus on the auxiliary 2D optical/Jacobian metrics used in the analysis. The study of null geodesics and LRs using the optical metric formalism was already well known in the literature. Using this method we recovered previous theorems concerning the number and stability of LRs in this class of spacetimes, albeit in a more restrictive case and with a different approach. For instance, one of these results states that BHs satisfying the above spacetime assumptions always have at least one unstable LR, with further LRs always coming in pairs with opposite stability. The analysis was then trivially extended to horizonless spacetimes, for which LRs can only come in pairs with opposing stability, with the innermost (outermost) being stable (unstable). 
	
	The LR analysis achieved through the optical metric approach serves additionally as an introduction to the similar study concerning TCOs, but now using the Jacobi rather than the optical metric. To the best of our knowledge, a detailed study of TCOs obtained using the Jacobi metric formalism (and MSCOs/ISCO in particular) constitutes a new contribution to the literature. The latter recovers some recent results concerning  the existence of TCOs and LRs within a spacetime. For instance, it is shown that no TCOs are possible radially above (bellow) a stable (unstable) LR. In addition, a novel theorem was proposed concerning the possible number of MSCOs, as well as their location and stability, in regions where TCOs are possible.

\section{Acknowledgements}

This work was supported by the Center for Research and Development in Mathematics and Applications (CIDMA) through the Portuguese Foundation for Science and Technology (FCT - Fundação para a Ci\^encia e a Tecnologia), references UIDB/04106/2020 and UIDP/04106/2020, by national funds (OE), through FCT, I.P., in the scope of the framework contract foreseen in the numbers 4, 5 and 6 of the article 23, of the Decree-Law 57/2016, of August 29, changed by Law 57/2017, of July 19 and by the projects PTDC/FIS-OUT/28407/2017,  CERN/FIS-PAR/0027/2019, PTDC/FIS-AST/3041/2020 and CERN/FIS-PAR/0024/2021. This work has further been supported by  the  European  Union's  Horizon  2020  research  and  innovation  (RISE) programme H2020-MSCA-RISE-2017 Grant No.~FunFiCO-777740 and by FCT through Project~No.~UIDB/00099/2020. PC is supported by the Individual CEEC program 2020 funded by the FCT.  J. P.A. Novo is supported by the FCT grant 2021.06539.BD.

	\bibliographystyle{unsrt}
	\bibliography{refsCurv}

\begin{thebibliography}{10}

\bibitem{Gibbons:2008rj}
G.~W. Gibbons and M.~C. Werner.
\newblock {Applications of the Gauss-Bonnet theorem to gravitational lensing}.
\newblock {\em Class. Quant. Grav.}, 25:235009, 2008.

\bibitem{Werner:2012rc}
M.~C. Werner.
\newblock {Gravitational lensing in the Kerr-Randers optical geometry}.
\newblock {\em Gen. Rel. Grav.}, 44:3047--3057, 2012.

\bibitem{Ono:2017pie}
Toshiaki Ono, Asahi Ishihara, and Hideki Asada.
\newblock {Gravitomagnetic bending angle of light with finite-distance
  corrections in stationary axisymmetric spacetimes}.
\newblock {\em Phys. Rev. D}, 96(10):104037, 2017.

\bibitem{Ishihara:2016vdc}
Asahi Ishihara, Yusuke Suzuki, Toshiaki Ono, Takao Kitamura, and Hideki Asada.
\newblock {Gravitational bending angle of light for finite distance and the
  Gauss-Bonnet theorem}.
\newblock {\em Phys. Rev. D}, 94(8):084015, 2016.

\bibitem{example1Jusufi:2015laa}
Kimet Jusufi.
\newblock {Gravitational lensing by Reissner-Nordstr\"om black holes with
  topological defects}.
\newblock {\em Astrophys. Space Sci.}, 361(1):24, 2016.

\bibitem{example2Ovgun:2018prw}
Ali \"Ovg\"un, Galin Gyulchev, and Kimet Jusufi.
\newblock {Weak Gravitational lensing by phantom black holes and phantom
  wormholes using the Gauss-Bonnet theorem}.
\newblock {\em Annals Phys.}, 406:152--172, 2019.

\bibitem{example3Ovgun:2018ran}
Ali Ovg\"un, Kimet Jusufi, and Izzet Sakalli.
\newblock {Gravitational lensing under the effect of Weyl and bumblebee
  gravities: Applications of Gauss\textendash{}Bonnet theorem}.
\newblock {\em Annals Phys.}, 399:193--203, 2018.

\bibitem{example4Ovgun:2018fnk}
Ali \"Ovg\"un.
\newblock {Light deflection by Damour-Solodukhin wormholes and Gauss-Bonnet
  theorem}.
\newblock {\em Phys. Rev. D}, 98(4):044033, 2018.

\bibitem{example5Jusufi:2018jof}
Kimet Jusufi, Ali \"Ovg\"un, Joel Saavedra, Yerko V\'asquez, and P.~A.
  Gonz\'alez.
\newblock {Deflection of light by rotating regular black holes using the
  Gauss-Bonnet theorem}.
\newblock {\em Phys. Rev. D}, 97(12):124024, 2018.

\bibitem{example6Zhu:2019ura}
Tao Zhu, Qiang Wu, Mubasher Jamil, and Kimet Jusufi.
\newblock {Shadows and deflection angle of charged and slowly rotating black
  holes in Einstein-\AE{}ther theory}.
\newblock {\em Phys. Rev. D}, 100(4):044055, 2019.

\bibitem{example7Kumar:2019ohr}
Rahul Kumar, Balendra~Pratap Singh, and Sushant~G. Ghosh.
\newblock {Shadow and deflection angle of rotating black hole in asymptotically
  safe gravity}.
\newblock {\em Annals Phys.}, 420:168252, 2020.

\bibitem{Islam:2020xmy}
Shafqat~Ul Islam, Rahul Kumar, and Sushant~G. Ghosh.
\newblock {Gravitational lensing by black holes in the $4D$
  Einstein-Gauss-Bonnet gravity}.
\newblock {\em JCAP}, 09:030, 2020.

\bibitem{Gibbons:2015qja}
G.~W. Gibbons.
\newblock {The Jacobi-metric for timelike geodesics in static spacetimes}.
\newblock {\em Class. Quant. Grav.}, 33(2):025004, 2016.

\bibitem{Chanda:2019guf}
Sumanto Chanda, G.~W. Gibbons, Partha Guha, Paolo Maraner, and Marcus~C.
  Werner.
\newblock {Jacobi-Maupertuis Randers-Finsler metric for curved spaces and the
  gravitational magnetoelectric effect}.
\newblock {\em J. Math. Phys.}, 60(12):122501, 2019.

\bibitem{Das:2016opi}
Praloy Das, Ripon Sk, and Subir Ghosh.
\newblock {Motion of charged particle in Reissner\textendash{}Nordstr\"om
  spacetime: a Jacobi-metric approach}.
\newblock {\em Eur. Phys. J. C}, 77(11):735, 2017.

\bibitem{Crisnejo:2018uyn}
Gabriel Crisnejo and Emanuel Gallo.
\newblock {Weak lensing in a plasma medium and gravitational deflection of
  massive particles using the Gauss-Bonnet theorem. A unified treatment}.
\newblock {\em Phys. Rev. D}, 97(12):124016, 2018.

\bibitem{Li:2019qyb}
Zonghai Li and Junji Jia.
\newblock {The finite-distance gravitational deflection of massive particles in
  stationary spacetime: a Jacobi metric approach}.
\newblock {\em Eur. Phys. J. C}, 80(2):157, 2020.

\bibitem{Delgado:2021jxd}
Jorge F.~M. Delgado, Carlos A.~R. Herdeiro, and Eugen Radu.
\newblock {Equatorial timelike circular orbits around generic ultracompact
  objects}.
\newblock {\em Phys. Rev. D}, 105(6):064026, 2022.

\bibitem{Barausse_2020}
Enrico Barausse, Emanuele Berti, Thomas Hertog, Scott~A. Hughes, Philippe
  Jetzer, Paolo Pani, Thomas~P. Sotiriou, Nicola Tamanini, Helvi Witek, Kent
  Yagi, Nicol{\'{a}}s Yunes, T.~Abdelsalhin, A.~Achucarro, K.~van Aelst,
  N.~Afshordi, S.~Akcay, L.~Annulli, K.~G. Arun, I.~Ayuso, V.~Baibhav,
  T.~Baker, H.~Bantilan, T.~Barreiro, C.~Barrera-Hinojosa, N.~Bartolo,
  D.~Baumann, E.~Belgacem, E.~Bellini, N.~Bellomo, I.~Ben-Dayan, I.~Bena,
  R.~Benkel, E.~Bergshoefs, L.~Bernard, S.~Bernuzzi, D.~Bertacca, M.~Besancon,
  F.~Beutler, F.~Beyer, S.~Bhagwat, J.~Bicak, S.~Biondini, S.~Bize, D.~Blas,
  C.~Boehmer, K.~Boller, B.~Bonga, C.~Bonvin, P.~Bosso, G.~Bozzola, P.~Brax,
  M.~Breitbach, R.~Brito, M.~Bruni, B.~Brügmann, H.~Bulten, A.~Buonanno, L.~M.
  Burko, C.~Burrage, F.~Cabral, G.~Calcagni, C.~Caprini,
  A.~C{\'{a}}rdenas-Avenda{\~{n}}o, M.~Celoria, K.~Chatziioannou, D.~Chernoff,
  K.~Clough, A.~Coates, D.~Comelli, G.~Comp{\`{e}}re, D.~Croon, D.~Cruces,
  G.~Cusin, C.~Dalang, U.~Danielsson, S.~Das, S.~Datta, J.~de~Boer, V.~De Luca,
  C.~De Rham, V.~Desjacques, K.~Destounis, F.~Di Filippo, A.~Dima,
  E.~Dimastrogiovanni, S.~Dolan, D.~Doneva, F.~Duque, R.~Durrer, W.~East,
  R.~Easther, M.~Elley, J.~R. Ellis, R.~Emparan, J.~M. Ezquiaga, M.~Fairbairn,
  S.~Fairhurst, H.~F. Farmer, M.~R. Fasiello, V.~Ferrari, P.~G. Ferreira,
  G.~Ficarra, P.~Figueras, S.~Fisenko, S.~Foffa, N.~Franchini, G.~Franciolini,
  K.~Fransen, J.~Frauendiener, N.~Frusciante, R.~Fujita, J.~Gair, A.~Ganz,
  P.~Garcia, J.~Garcia-Bellido, J.~Garriga, R.~Geiger, C.~Geng, L.~{\'{A}}.
  Gergely, C.~Germani, D.~Gerosa, S.~B. Giddings, E.~Gourgoulhon,
  P.~Grandclement, L.~Graziani, L.~Gualtieri, D.~Haggard, S.~Haino, R.~Halburd,
  W.-B. Han, A.~J. Hawken, A.~Hees, I.~S. Heng, J.~Hennig, C.~Herdeiro,
  S.~Hervik, J.~v.~Holten, C.~J.~D. Hoyle, Y.~Hu, M.~Hull, T.~Ikeda, M.~Isi,
  A.~Jenkins, F.~Juli{\'{e}}, E.~Kajfasz, C.~Kalaghatgi, N.~Kaloper,
  M.~Kamionkowski, V.~Karas, S.~Kastha, Z.~Keresztes, L.~Kidder, T.~Kimpson,
  A.~Klein, S.~Klioner, K.~Kokkotas, H.~Kolesova, S.~Kolkowitz, J.~Kopp,
  K.~Koyama, N.~V. Krishnendu, J.~A.~V. Kroon, M.~Kunz, O.~Lahav, A.~Landragin,
  R.~N. Lang, C.~Le Poncin-Lafitte, J.~Lemos, B.~Li, S.~Liberati, M.~Liguori,
  F.~Lin, G.~Liu, F.~S.~N. Lobo, R.~Loll, L.~Lombriser, G.~Lovelace, R.~P.
  Macedo, E.~Madge, E.~Maggio, M.~Maggiore, S.~Marassi, P.~Marcoccia,
  C.~Markakis, W.~Martens, K.~Martinovic, C.~J. A.~P. Martins, A.~Maselli,
  S.~Mastrogiovanni, S.~Matarrese, A.~Matas, N.~E. Mavromatos, A.~Mazumdar,
  P.~D. Meerburg, E.~Megias, J.~Miller, J.~P. Mimoso, L.~Mittnacht, M.~M.
  Montero, B.~Moore, P.~Martin-Moruno, I.~Musco, H.~Nakano, S.~Nampalliwar,
  G.~Nardini, A.~Nielsen, J.~Nov{\'{a}}k, N.~J. Nunes, M.~Okounkova,
  R.~Oliveri, F.~Oppizzi, G.~Orlando, N.~Oshita, G.~Pappas, V.~Paschalidis,
  H.~Peiris, M.~Peloso, S.~Perkins, V.~Pettorino, I.~Pikovski, L.~Pilo,
  J.~Podolsky, A.~Pontzen, S.~Prabhat, G.~Pratten, T.~Prokopec, M.~Prouza,
  H.~Qi, A.~Raccanelli, A.~Rajantie, L.~Randall, G.~Raposo, V.~Raymond,
  S.~Renaux-Petel, A.~Ricciardone, A.~Riotto, T.~Robson, D.~Roest, R.~Rollo,
  S.~Rosofsky, J.~J. Ruan, D.~Rubiera-Garc{\'{\i}}a, M.~Ruiz, M.~Rusu,
  F.~Sabatie, N.~Sago, M.~Sakellariadou, I.~D. Saltas, L.~Sberna,
  B.~Sathyaprakash, M.~Scheel, P.~Schmidt, B.~Schutz, P.~Schwaller, L.~Shao,
  S.~L. Shapiro, D.~Shoemaker, A.~d.~Silva, C.~Simpson, C.~F. Sopuerta,
  A.~Spallicci, B.~A. Stefanek, L.~Stein, N.~Stergioulas, M.~Stott, P.~Sutton,
  R.~Svarc, H.~Tagoshi, T.~Tahamtan, H.~Takeda, T.~Tanaka, G.~Tantilian,
  G.~Tasinato, O.~Tattersall, S.~Teukolsky, A.~L. Tiec, G.~Theureau,
  M.~Trodden, A.~Tolley, A.~Toubiana, D.~Traykova, A.~Tsokaros, C.~Unal, C.~S.
  Unnikrishnan, E.~C. Vagenas, P.~Valageas, M.~Vallisneri, J.~Van den Brand,
  C.~Van den Broeck, M.~van~de Meent, P.~Vanhove, V.~Varma, J.~Veitch,
  B.~Vercnocke, L.~Verde, D.~Vernieri, F.~Vernizzi, R.~Vicente, F.~Vidotto,
  M.~Visser, Z.~Vlah, S.~Vretinaris, S.~Völkel, Q.~Wang, Yu-Tong Wang, M.~C.
  Werner, J.~Westernacher, R.~v.~d. Weygaert, D.~Wiltshire, T.~Wiseman,
  P.~Wolf, K.~Wu, K.~Yamada, H.~Yang, L.~Yi, X.~Yue, D.~Yvon, M.~Zilh{\~{a}}o,
  A.~Zimmerman, and M.~Zumalacarregui.
\newblock Prospects for fundamental physics with {LISA}.
\newblock {\em General Relativity and Gravitation}, 52(8), aug 2020.

\bibitem{Qiao:2022jlu}
Chen-Kai Qiao and Ming Li.
\newblock {A Geometric Approach on Circular Photon Orbits and Black Hole
  Shadow}.
\newblock 4 2022.

\bibitem{Cvetic:2016bxi}
M.~Cvetic, G.~W. Gibbons, and C.~N. Pope.
\newblock {Photon Spheres and Sonic Horizons in Black Holes from Supergravity
  and Other Theories}.
\newblock {\em Phys. Rev. D}, 94(10):106005, 2016.

\bibitem{Cunha:2017qtt}
Pedro V.~P. Cunha, Emanuele Berti, and Carlos A.~R. Herdeiro.
\newblock {Light-Ring Stability for Ultracompact Objects}.
\newblock {\em Phys. Rev. Lett.}, 119(25):251102, 2017.

\bibitem{Cunha:2020azh}
Pedro V.~P. Cunha and Carlos A.~R. Herdeiro.
\newblock {Stationary black holes and light rings}.
\newblock {\em Phys. Rev. Lett.}, 124(18):181101, 2020.

\bibitem{Carter:1973rla}
B.~Carter.
\newblock {Black holes equilibrium states}.
\newblock In {\em {Les Houches Summer School of Theoretical Physics}: {Black
  Holes}}, pages 57--214, 1973.

\bibitem{Wald:1999vt}
Robert~M. Wald.
\newblock {The thermodynamics of black holes}.
\newblock {\em Living Rev. Rel.}, 4:6, 2001.

\bibitem{Wald:1984rg}
Robert~M. Wald.
\newblock {\em {General Relativity}}.
\newblock Chicago Univ. Pr., Chicago, USA, 1984.

\bibitem{Brito:2015pxa}
Richard Brito, Vitor Cardoso, Carlos A.~R. Herdeiro, and Eugen Radu.
\newblock {Proca stars: Gravitating Bose\textendash{}Einstein condensates of
  massive spin 1 particles}.
\newblock {\em Phys. Lett. B}, 752:291--295, 2016.

\bibitem{Keir:2014oka}
Joe Keir.
\newblock {Slowly decaying waves on spherically symmetric spacetimes and
  ultracompact neutron stars}.
\newblock {\em Class. Quant. Grav.}, 33(13):135009, 2016.

\bibitem{Cardoso:2014sna}
Vitor Cardoso, Lu\'\i{}s C.~B. Crispino, Caio F.~B. Macedo, Hirotada Okawa, and
  Paolo Pani.
\newblock {Light rings as observational evidence for event horizons: long-lived
  modes, ergoregions and nonlinear instabilities of ultracompact objects}.
\newblock {\em Phys. Rev. D}, 90(4):044069, 2014.

\bibitem{Cunha:2022gde}
Pedro V.~P. Cunha, Carlos Herdeiro, Eugen Radu, and Nicolas Sanchis-Gual.
\newblock {The fate of the light-ring instability}.
\newblock 7 2022.

\bibitem{Carroll:2004st}
Sean~M. Carroll.
\newblock {\em {Spacetime and Geometry}}.
\newblock Cambridge University Press, 7 2019.

\bibitem{Wei:2022mzv}
Shao-Wen Wei and Yu-Xiao Liu.
\newblock {"Topology of equatorial timelike circular orbits around stationary
  black holes", arXiv: gr-qc/2207.08397}.
\newblock (2022).

\bibitem{Hawking:1971vc}
S.~W. Hawking.
\newblock {Black holes in general relativity}.
\newblock {\em Commun. Math. Phys.}, 25:152--166, 1972.

\bibitem{Hawking:1973uf}
S.~W. Hawking and G.~F.~R. Ellis.
\newblock {\em {The Large Scale Structure of Space-Time}}.
\newblock Cambridge Monographs on Mathematical Physics. Cambridge University
  Press, 2 2011.

\bibitem{docarmo}
Manfredo~P. Do~Carmo.
\newblock {\em Differential geometry of curves \& surfaces}.
\newblock Dover Publications, Inc., 2016.

\bibitem{Note1}
In fact, Hawking showed that assuming that the dominant energy condition is
  obeyed the event horizon of $4$-dimensional asymptotically flat stationery
  black holes is always topologically spherical\cite
  {Hawking:1971vc,Hawking:1973uf}.

\bibitem{Thornburg:2006zb}
Jonathan Thornburg.
\newblock {Event and apparent horizon finders for 3+1 numerical relativity}.
\newblock {\em Living Rev. Rel.}, 10:3, 2007.

\end{thebibliography}
	
	\pagebreak
	
	\appendix
	
	\section{Geometry of surfaces \label{sec:Surfaces}}
	
	In this appendix the relevant quantities relative to the geometry of 2 dimensional surface will be introduced and computed. Namely the Gaussian curvature of the manifold and the geodesic curvatures of circular orbits.
	
	To do so a generic 2 dimensional manifold, $\mathcal{S}$, equipped with a Riemannian metric, $\alpha$. This can represent both the optical or Jacobi manifold. The line element on such surface is therefore given by:
	\begin{equation}
		{\rm d}\lambda^{2}=\alpha_{rr}\left(r\right){\rm d}r^{2}+\alpha_{\phi\phi}\left(r\right){\rm d}\phi^{2}\,.\label{eq:2lineele}
	\end{equation}
	Here $\lambda$ will be the coordinate time $t$ when considering the optical metric, and the Jacobi parameter $s_J$ when discussing the Jacobi metric. It was assumed that the metric components depend only on the radial coordinates, as this will be the case for the metrics considered in this paper.
	
	\subsection{Gaussian curvature}
	
	Considering the form of the metric, Eq. (\ref{eq:2lineele}), it is useful to introduce the Regge-Wheeler tortoise coordinate, $r^*$, defined as
	\begin{equation}
		\mathrm{d}r^*=\sqrt{\alpha_{rr}}\mathrm{d}r\,.
	\end{equation}
	In these coordinates the line element becomes
	\begin{equation}
		{\rm d}\lambda^{2}={\rm d}r^{*2}+\alpha_{\phi\phi}\left(r\left(r^*\right)\right){\rm d}\phi^{2}\,.
	\end{equation}
	This makes evident that $\mathcal{S}$ is a surface of revolution, thus its Gaussian curvature is given by
	\begin{equation}
		K=-\frac{1}{\alpha_{\phi\phi}\left(r\left(r^*\right)\right)}\frac{\mathrm{d}^2  \alpha_{\phi\phi}\left(r\left(r^*\right)\right)}{\mathrm{d}r^{*2}}\,.
	\end{equation}
	
	This set of coordinates is also usually called \emph{geodesic polar coordinates} \cite{docarmo}, which are useful to understand the Gaussian curvature. These correspond to polar coordinates in the tangent plane $T_p\left(S\right)$ at a given point $p\in S$. This system is usually defined by the polar radius, $r^*$, and a polar angle, $\phi\in ]0,2\pi[$, this coordinate system has a pole at the origin of $T_p\left(S\right)$. In these coordinates the metric components obey:
	\begin{equation}
		\alpha_{r^*r^*}=1\,,\: \alpha_{r^*\phi}=0\,,\: \lim_{r^*\rightarrow0} \alpha_{\phi\phi}=0\,,\: \lim_{r^*\rightarrow0}\frac{\partial\sqrt{\alpha_{\phi\phi}}}{\partial r^*}=1\,.
	\end{equation}
	This allows for an intuitive interpretation of the Gaussian curvature. The arc-length of the curve with $r^*=\mathrm{const.}$ joining two close geodesics with $\vartheta=\vartheta_1$ and $\vartheta=\vartheta_2$, $L(\rho)$, is given by
	\begin{equation}
		L(r^*)=\intop_{\vartheta_1}^{\vartheta_2}\sqrt{\alpha_{\vartheta\vartheta}}\,\mathrm{d}\vartheta\,.
	\end{equation}
	Then, since 
	\begin{equation}
		\lim_{r^*\rightarrow0}\frac{\partial\sqrt{\alpha_{\phi\phi}}}{\partial r^*}=1\,,\quad \frac{\partial^2\sqrt{\alpha_{\phi\phi}}}{\partial r^{*2}}=-K \sqrt{\alpha_{\phi\phi}}\,,
	\end{equation}
	we have that if $K<0$ the distance between two geodesics starting close together will continue to grow. On the other hand, if $K>0$ the geodesics can begin to approach after some time.
	
	\subsection{Geodesic curvature of circular orbits}
	
	Consider a curve $\gamma$ on ${\cal S}$, with arc-length parameter $\lambda$ its geodesic curvature, $\kappa_{g}$, is given by
	\begin{equation}
		\kappa_{g}=\frac{{\rm d}\mathbf{T}}{{\rm d}\lambda}\cdot\mathbf{n}\,, \label{eq:GeoCurv}
	\end{equation}
	where $\mathbf{T}={\rm d}\mathbf{r}\left(\lambda\right)/{\rm d}\lambda$ is the 	unit tangent vector to $\gamma$ and $\mathbf{n}$ the unit normal vector to $ \gamma $. Therefore the geodesic curvature is the component of the proper acceleration along the normal direction to the curve. For circular geodesics, $r=R$, of the metric (\ref{eq:2lineele}) one has 
	\begin{equation}
		T^{i}=\delta_{\phi}^{i}\frac{{\rm d}\phi}{{\rm d}\lambda}=\delta_{\phi}^{i}\frac{1}{\sqrt{\alpha_{\phi\phi}}}\,.
	\end{equation}
	Since we are considering circular geodesics the unit normal vector to the curve can be $n_{i}=\sqrt{\alpha_{rr}}\delta_{i}^{r}$. Hence the geodesic curvature of circular curves is given by 
	\begin{align}
		\kappa_{g} & =\left.n_{i}\frac{{\rm d}T^{i}}{{\rm d}s}\right|_{r=R}=\left.n_{i}T^{j}\nabla_{j}T^{i}\right|_{r=R}\nonumber \\
		& =\left.n_{i}T^{j}\left(\partial_{j}T^{i}+\Gamma_{jl}^{i}T^{l}\right)\right|_{r=R}\nonumber \\
		& =\left.\delta_{i}^{r}\delta_{\phi}^{j}\delta_{\phi}^{l}\Gamma_{jl}^{i}\left(\frac{{\rm d}\phi}{{\rm d}\lambda}\right)^{2}\sqrt{\alpha_{rr}}\right|_{r=R}\nonumber \\
		& =-\left.\frac{1}{2\sqrt{\alpha_{rr}}}\frac{\partial\ln\left(\alpha_{\phi\phi}\right)}{\partial r}\right|_{r=R}\,.
	\end{align}
	
	It should be stressed that the sign of the geodesic curvature changes when either the orientation of the curve or of the surface changes. This corresponds to choosing the inwards pointing normal vector in our computations, $n_{i}=-\sqrt{\alpha_{rr}}\delta_{i}^{r}$. Therefore, the sign of $ \kappa_{g} $ simply indicates if the acceleration has the same or opposite sense as the chosen normal vector. For our choice it is negative, since circular orbits around compact objects have inwards pointing accelerations (particles are being pulled to the central object) and we chose the outwards pointing normal vector. Therefore only the absolute value of the geodesic curvature is intrinsic to the curve. Our choice was made in order to recover the asymptotic $ 1/r $ behaviour, characteristic of Euclidean geometry, for the optical and Jacobi manifolds.

	\section{Horizons \label{horizons}}
	
	BH spacetimes are characterised by the existence of an event horizon. This horizon corresponds to a hypersurface enclosing all 	points which are not connected to infinity by a timelike path. The notion of infinity may be very subtle when considering generic spacetimes, however for the present study of asymptotically flat spacetimes infinity corresponds precisely to this region where the spacetime is Minkowski. One can equivalently define the event horizon as the boundary of the closure of the causal past of future null infinity.
	
	This definition makes clear that the horizon is generated by null geodesics, hence it is a null hypersurface, $\Sigma$. Such hypersurface can be defined as a level curve of a given scalar function $f\left(x\right)$, 	where $x$ denotes the spacetime coordinates. The gradient of this function is normal to $\Sigma$, however, since null vectors are orthogonal to themselves the normal vector $\partial_{\mu}f$ is also tangent to $\Sigma$. This means that a hypersurface is generated by null 	geodesics. 
	
	This local definition relies on having an appropriate set of coordinates which is not always guaranteed. However, in this work we will be concerned only with static, asymptotically flat spacetimes which contain event horizons with spherical topology, such that it is possible to find a suitable coordinate system \cite{Carroll:2004st} \footnote{In fact, Hawking showed that assuming that the dominant energy condition is obeyed the event horizon of $4$-dimensional asymptotically flat stationery black holes is always topologically spherical\cite{Hawking:1971vc,Hawking:1973uf} }. These metrics possess a Killing vector field, $\partial_{t}$, which is
	asymptotically timelike, and the metric components can be adapted such that $\partial_{t}g_{\mu\nu}=0$. On hypersurfaces where $t=\mathrm{const.}$ it is possible to choose coordinates $\left(r,\theta,\phi\right)$
	in which the metric at infinity looks like Minkowski space in spherical polar coordinates. Hypersurfaces with $r=\mathrm{const}.$ will then be timelike cylinders with topology $S^{2}\times\mathbb{R}$ at $r\rightarrow\infty$. For a clever choice of such coordinates we can have the hypersurfaces 	$r=\mathrm{\mathrm{const.}}$ remain timelike from infinity all the way up to some $r_{H}$, for which the hypersurface is everywhere null. This will represent the event horizon, as timelike paths that cross into $r<r_{H}$ will never be able to escape back to infinity. Determining the radius at which the hypersurfaces with $r=\mathrm{const.}$ become null is simple, the normal one form normal to such hypersurfaces is $\partial_{\mu}r$, with norm
	\begin{equation}
		g^{\mu\nu}\partial_{\mu}r\partial_{\nu}r=g^{rr}\,.
	\end{equation}
	The null quality of such hypersurface implies that 
	\begin{equation}
		g^{rr}\left(r_{H}\right)=0\,.
	\end{equation}
	This is the condition that determines the location of the horizon for static, spherically symmetric, and asymptotically flat spacetimes \cite{Carroll:2004st,Thornburg:2006zb}.
	
	Another relevant kind of horizons are the Killing horizons, these correspond to hypersurfaces where a given Killing vector field is null. In the previous discussion about event horizons we made no mention of Killing
	horizons, however the two definitions are intimately connected by several theorems, known as rigidity theorems which state that under which conditions the event horizon is simultaneously a Killing horizon \cite{Carter:1973rla,Hawking:1973uf}. In fact for \emph{static} spacetimes the event horizon is simultaneously the  Killing horizon of the Killing vector field responsible for time translations at infinity \cite{Carter:1973rla}.
	
	
	\section{Surface gravity\label{sec:SurfGravity}}
	
	A null hypersurface ${\cal N}$ is a Killing horizon of a Killing vector field $\xi$ if, on ${\cal N}$, $\xi$ null. This Killing vector field will also be normal to $ \mathcal{N} $, since a null hypersurface cannot have two linearly independent null tangent vectors.
	
	Since at the horizon $\xi$ is null the hypersurface ${\cal N}$ is generated by some $k=f\xi$ also, where $f$ is some function and $ k^\mu\nabla_\mu k^\nu=0 $.
	
	The surface gravity $\kappa_{\mathrm{surf}}$ at such an horizon is defined as
	\begin{equation}
		\xi^{\alpha}\nabla_{\alpha}\xi^{\beta}=\kappa_{\mathrm{surf}}\xi^{\beta}\,.
	\end{equation}
	Therefore it is a measure on how orbits of $ \xi^\alpha $ fail to be geodesics. This is equivalent to:
	\begin{equation}
		\kappa_{\mathrm{surf}}^{2}=\left.\frac{1}{2}\nabla_{\alpha}\xi_{\beta}\nabla^{\alpha}\xi^{\beta}\right|_{\cal N}\,.
	\end{equation}
	For the present case we will consider the Killing vector field associated with time translations. Since our metric does not explicitly depend on the time coordinate we have that in our coordinate system $\xi^{\alpha}=\delta_{t}^{\alpha}$.
	Firs we compute $\nabla_{\alpha}\xi^{\beta}$
	\begin{align}
		\nabla_{\alpha}\xi^{\beta}= & \partial_{\alpha}\xi^{\beta}+\Gamma_{\alpha\lambda}^{\beta}\xi^{\lambda}\nonumber \\
		= & \Gamma_{\alpha t}^{\beta}\nonumber \\
		= & \left(-\frac{\delta_{\alpha}^{t}}{2}g^{\beta\sigma}\partial_{\sigma}g_{tt}+\delta_{t}^{\beta}\partial_{\alpha}\ln\sqrt{g_{tt}}\right)\nonumber \\
		= & -\frac{\delta_{\alpha}^{t}\delta_{r}^{\beta}}{2}\partial_{r}g_{tt}+\delta_{t}^{\beta}\delta_{\alpha}^{r}\partial_{r}\ln\sqrt{g_{tt}}\,.
	\end{align}
	Expanding the expression defining the surface gravity one obtains
	\begin{equation}
		\kappa_{\mathrm{surf}}^{2}=\left.\frac{1}{4}g^{tt}g_{rr}\left(g^{rr}\right)^{2}\left(\partial_{r}g_{tt}\right)^{2}+\frac{g^{rr}}{2}\partial_{r}g_{tt}\right|_{\cal N}\,.
	\end{equation}
	The event horizon is defined by the condition $g^{rr}\left(r_{H}\right)=0$,
	hence the surface gravity at the horizon is given by
	\begin{equation}
		\kappa_{\mathrm{surf}}=\lim_{r\rightarrow r_{H}}\frac{1}{2}\sqrt{\frac{h\left(r\right)}{f\left(r\right)}}f^\prime\left(r\right)\,.\label{eq:SurfaceSph}
	\end{equation}
	It should be noted that if $ \mathcal{N} $ is a Killing horizon of $ \xi $ it will also be a Killing horizon of some $ c \xi $, where $ c $ is some constant, with surface gravity $ c^2 \kappa_{\mathrm{surf}} $. This means that the surface gravity is not an intrinsic property of $ \mathcal{N} $, but also depends on the normalisation of $ \xi $. At the horizon  $ \xi $ is null, hence it does not admit any natural normalisation there. However, for asymptotically flat spacetimes it can be normalised such that:
	\begin{equation}
		\lim_{r\rightarrow \infty} \xi_{\beta}\xi^\beta=-1\,.
	\end{equation}
	This fixes $ \kappa_{\mathrm{surf}} $ up to a sign, which is further fixed by requiring $ \xi $ to be future directed.
	
	Expression (\ref{eq:SurfaceSph}) reveals a maybe unexpected relation between the surface gravity and the geodesic curvature, this is by no means fortuitous. For spherically symmetric BHs the surface gravity, with the above normalisation, is the acceleration of a static observer at the horizon, as measured by a static observer at infinity. Therefore it is not surprising that the surface gravity corresponds to the geodesic curvature of the circular null generators of the event horizon, apart from a sign coming from a choice of orientation of the curve.
	
	\section{Potential approach \label{sec:AppPotentialApproach}}
	
	The motion of test particles in general relativity can be obtained from the following effective Lagrangian:
	\begin{equation}
		2{\cal L}=g_{\mu\nu}\dot{x}^{\mu}\dot{x}^{\nu}=\sigma\,,
	\end{equation}
	where  $\sigma=-1,0,+1$ for timelike, null and spacelike orbits, respectively. Using the line element \ref{eq:LineElement} this can be cast in the following form
	\begin{equation}
		g_{rr}\dot{r}^{2}=\sigma-\frac{f\left(r\right)L^{2}-\varepsilon^{2}}{r^{2}f\left(r\right)}\equiv V_{eff}\,,
	\end{equation}
	where the effective potential $ V_{eff} $ was introduced. Circular geodesics must have
	\begin{equation}
		\dot{r}=0\,\wedge\ddot{r}=0\Rightarrow V_{eff}=0\,\wedge V^\prime_{eff}=0\,.\label{eq:VeffConditions}
	\end{equation}
	
	For null geodesics ($\sigma=0$) the circular null geodesics are obtained then by 
	\begin{equation}
		V_{eff}^{\prime}=\frac{\varepsilon}{f\left(r\right)^{2}}\frac{2f\left(r\right)-rf^{\prime}\left(r\right)}{r}=0\,.
	\end{equation}
	The roots of this equation are same of Eq. (\ref{eq:kappaLR}), attesting the equivalence between the two approaches.
	
	For timelike geodesics ($\xi=-1$) we will be concerned with the MSCOs. At a linear level the stability of the TCOs is determined by the sign of $ V_{eff}^{\prime\prime} $, hence one must add the condition $ V_{eff}^{\prime\prime} $ to the ones on Eq. (\ref{eq:VeffConditions}). This yields:
	\begin{equation}
		V_{eff}^{\prime\prime}=2\frac{f(r)\left(3f'(r)+rf''(r)\right)-2rf'(r)^{2}}{rf(r)\left(rf'(r)-2f(r)\right)}\,.
	\end{equation}
	Once more the roots of this expression coincide with the ones of Eq. (\ref{eq:GaussMSCO}), meaning that our approach yields the same results as the usual one. A curious note is that this expression diverges at the LRs while Eq. (\ref{eq:GaussMSCO}) vanishes.

\end{document}